# Translation position extracting in incoherent Fourier ptychography

ZONGLIANG XIE,[1,2,3,4] HAOTONG MA,[2,3,4,5] YIHAN LUO,[2,3,4,6] BO QI,[2,3,4] AND GE REN[2,3,4]

[1]State Key Laboratory of Pulsed Power Laser Technology, Hefei, 230037, China
[2]Key Laboratory of Optical Engineering, Chinese Academy of Sciences, Chengdu, 610209, China
[3]Institute of Optics and Electronics, Chinese Academy of Sciences, Chengdu, 610209, China
[4]University of Chinese Academy of Sciences, Beijing, 100039, China
[5]mahaotong@163.com
[6]luo.yihan@foxmail.com

**Abstract:** Incoherent Fourier ptychography (IFP) is a newly developed super-resolution method, where accurate knowledge of translation positions is essential for image reconstruction. To release this limitation, we propose a preprocessing algorithm capable of extracting translation positions of the structure light directly from raw images of IFP, termed translation position extracting (TPE). TPE mainly involves two steps. First, the speckle parts mixed in the acquired intensities, in which the illumination motion is encoded, are isolated by intensity averaging and division. Then the cross-correlations of the speckle dataset are computed to determine the shift positions. TPE-IFP improves the previous IFP by removal of the requirement for prior knowledge of translation positions. Its effectiveness is demonstrated by obtaining high-quality super-resolution images in absence of location information in both simulations and experiments. By further relaxing the practical conditions, the proposed TPE may accelerate the applications of IFP. What's more, as a preprocessing approach, TPE might also contribute to the estimation of pattern positions for the similar speckle-based imaging.

## 1. Introduction

Incoherent Fourier ptychography (IFP) is a super-resolution imaging technique using structured light for the incoherent case, where the spatial high-frequency can be shifted to the passband of the imaging system by frequency mixing between the object and the illumination speckle. Sharing its roots with Fourier ptychography [1-3], the IFP algorithm decodes the high-frequency by iteratively switching between the spatial and Fourier domains. IFP has shown the potential in fluorescence microscopy [4], macroscopic photography [5], optical synthetic aperture [6] and multi-dimension imaging [7]. It also motivates other useful supper-resolution imaging methods via translated speckle illumination, such as fluorescence imaging through a turbid layer [8] and near-field Fourier ptychography [9].

In the recovery process of IFP, the specific distribution of the illumination pattern isn't required while the accurate knowledge of its translation positions relative to the object needs to be known. In complex circumstances, such as the case where the object moves quickly, acquiring the prior position information seems challenging. A pattern-estimation algorithm using the gradient descent framework for this drawback has been reported [10]. Though successfully eliminating the requirement of the illumination knowledge, this method needs to repeat the nonlinear optimization process serially for all intensity measurements in one iteration step of IFP, thus imposing a heavy load on computational resources.

In this paper, we propose a simple and efficient preprocessing approach for IFP, termed translation position extracting (TPE), which is able to extract relative translation positions of the structure light directly from raw images. The captured intensity can be modeled as a product of two components: object and speckle distributions. Their relative movement could be considered as that the speckle parts move with the illumination translation while the object parts remain constant. In this case, the motion is encoded in speckle distribution. Based on IFP raw dataset, TPE first isolates the pattern by means of intensity averaging and division, and then utilizes the cross-correlation to estimate the illumination motion. Embedded with TPE, IFP can reconstruct a high-quality supper-resolution image without knowing the prior translation positions. We also note that as a preprocessing method, TPE acquires the speckle movement from the raw dataset while being irrelevant with the image reconstruction algorithm, and thus is possible to determine the pattern translation positions for the similar speckle-based supper-resolution imaging techniques [8,9].

This paper is structured as follows. In section 2, the working principle of IFP is briefly reviewed and TPE algorithm we use to determine the positions is described in detail. In Section 3, we verify the effectiveness of our proposed TPE-IFP and analyze the accuracy of position estimation by simulation. Section 4 presents the experimental demonstration that the reported TPE-TFP can recover a high-quality image without the prior motion information. Finally, this paper is concluded in section 5.

## 2. Method

### 2.1 Incoherent Fourier ptychography

Relying on the hardware with scanning structured light, IFP uses a reconstruction algorithm stitching between spatial and Fourier domains to realize super-resolution imaging. Specifically, the illumination speckle is scanned in two dimensions, projecting an unknown random pattern onto the object, and correspondingly raw images modulated by the moving pattern are captured. In this way, the frequency between the object and speckle is mixed so that the information beyond the cutoff frequency is contained. Based on the acquired raw images, iteratively updating the spatial and Fourier domains by the similar model of the hybrid input-output algorithm can decode the hidden high-frequency and outputs a super-resolution image. IFP algorithm can be referred in detail in [4], and here is briefly reviewed as follows:

1. Start with initial guesses of the object intensity $I_{obj,0}$ and the unknown illumination pattern $P_0$. The optical transfer function $OTF$ of the imaging system, which could be either a single [5] or multi-aperture [6] system, is known, serving as a support constrain in Fourier domain. If necessary, thanks to the prior work [6,10], $OTF$ can also be updated with the reconstruction proceeding.

2. For the $n$th iteration, generate a target image as follows:

$$I_{tn} = I_{obj,n-1} \cdot P_{n-1}(\mathbf{v} - \mathbf{v}_n) \tag{1}$$

where $\mathbf{v}_n$ ($n = 1, 2, 3…$) presents the motion vector of the structured light relative to the object, which is essential knowledge for image reconstruction. The speckle motion vector $\mathbf{v}_n = (x_n, y_n)$ is defined as translating the pattern horizontally by $x_n$, and vertically by $y_n$, while the object remains constant.

3. Apply the captured image $I_n$ in the $n$th position to obtain an updated target image in the Fourier domain

$$\mathcal{F}\left(I_{tn}^{update}\right) = \mathcal{F}\left(I_{tn}\right) + OTF \cdot [\mathcal{F}\left(I_n\right) - OTF \cdot \mathcal{F}\left(I_{tn}\right)] \tag{2}$$

where $\mathcal{F}(\cdot)$ denotes Fourier transform.

4. Renew the object image in the spatial domain by using the updated target image $I_{tn}^{update}$ and projection pattern $P_{n-1}$ according to

$$I_{obj,n}^{update} = I_{obj,n-1} + \frac{P_{n-1}(\mathbf{v}-\mathbf{v}_n)}{\left(\max(P_{n-1}(\mathbf{v}-\mathbf{v}_n))\right)^2} \cdot \left(I_{tn}^{update} - I_{obj,n-1} \cdot P_{n-1}(\mathbf{v}-\mathbf{v}_n)\right) \quad (3)$$

5. Update the illumination speckle with the following equation

$$P_n(\mathbf{v}-\mathbf{v}_n) = P_{n-1}(\mathbf{v}-\mathbf{v}_n) + \frac{I_{obj,n}^{update}}{\left(\max(I_{obj,n}^{update})\right)^2} \cdot \left(I_{tn}^{update} - I_{tn}\right) \quad (4)$$

The steps 2 to 5 described above repeats for all the captured images. The whole process continues until the recovery is converged.

### 2.2 Translation position extracting algorithm

Before IFP reconstruction procedure, TPE is proposed to acquire the important motion information directly from the raw images. The object is modulated by the illumination pattern, resulting in captured images coupling the object and speckle distributions. The essential step to extract the translation positions is isolating the speckle pattern, where the motion is encoded. Here we establish simple mathematic models with averaging and division to perform that important operation, which seems challenging at first glance.

First, we compute the pixel-wise mean image $I^{mean}$ of the raw image sequence $[I_1, I_2, ..., I_n]$ as follows:

$$I^{mean} = \frac{1}{N}\sum_{n=1}^{N} I_n \quad (5)$$

where $N$ is the total number of the raw images. Owing to the statistical randomness, the speckle components approximately get averaged out of the $I^{mean}$. That is to say, $I^{mean}$ can be regarded as the object part in the captured images. Hence, approximatively the pattern component can be isolated by computing the ratio image

$$I_n^{speckle}(\mathbf{v}-\mathbf{v}_n) \approx \frac{I_n}{I^{mean}} \quad (6)$$

where $I_n^{speckle}(\mathbf{v}-\mathbf{v}_n)$ is the speckle part of the captured image corresponding to the translation position $\mathbf{v}_n$, in which the motion is encoded.

Then the cross-correlation algorithm is used to determine the positions, as motivated in [11]. We compute the cross-correlation between each speckle component and a constant one taken as the reference in sequence following

$$C_n(\mathbf{v}) = I_n^{speckle}(\mathbf{v}-\mathbf{v}_n) \otimes I_{refer}^{speckle}(\mathbf{v}) \quad (7)$$

where $I_{refer}^{speckle}(\mathbf{v})$ is the reference speckle, which can be chosen randomly. In IFP, there are large overlapped regions among all the speckle images, so the cross-correlation can be approximately rewritten as

$$C_n(\mathbf{v}) \approx \left[I_{refer}^{speckle}(\mathbf{v}) \otimes I_{refer}^{speckle}(\mathbf{v})\right] * \delta(\mathbf{v}-\mathbf{v}_n) \quad (8)$$

where $*$ represents the convolution and $\delta(\cdot)$ is the Dirac delta function. In this way, the shift vector $\mathbf{v}_n$ can be extracted by locating the cross-correlation peak.

### 3. Simulation analysis

To check the validity of the proposed TPE and evaluate its performance, we perform a series of simulations. Simulations are conducted based on an imaging system with the diameter of 10 mm and the focal length of 300 mm. The pixel size of the detector is 3.45 μm and the imaging wavelength is 632 nm. We use a picture shown in Fig. 1(a) as the object. A fully randomized intensity distribution shown as Fig. 1(b) is used as the illumination pattern. In the simulation, we scan the speckle pattern to illuminate the object. Specifically, the pattern is shifted to 81 positions spaced equally in 2 dimensional square grid with the interval of 10 pixels. At each position, an image coupling the object and speckle distributions is captured. To increase the reality of the simulation, Gaussian noises (noise variance $\sigma^2 = 0.1\%$ of the variance of signal values) are added. 4 representations of the raw images are shown as Figs. 1(c1)-4(c4).

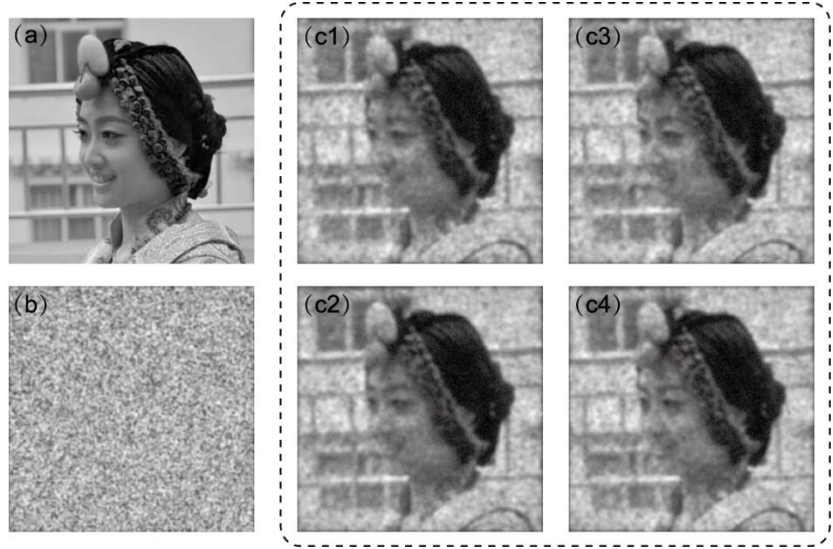

Fig. 1. Simulated raw images modulated by the translated speckle. (a) The object and (b) the illumination speckle used in the simulation. (c1)-(c4) Four frames of the captured raw images.

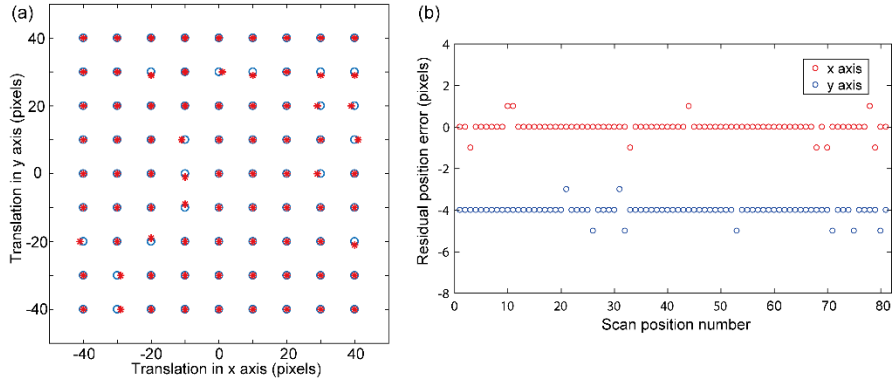

Fig. 2. Simulation results of the estimation of the pattern translation. (a) The extracted positions marked with asterisks and the ground-truths marked with the circles. (b) The distribution of residual position errors.

Accurate knowledge of translation positions is essential for IFP to reconstruct high-quality images. Luckily, TPE can bypass this limitation by extracting the encoded motion directly from the raw images. The speckle components are isolated by averaging and division. Based on the pattern dataset, the cross-correlation technique is utilized to determine the translation positions. Figure 2(a) presents the results of position estimation, where the circles represent the ground-truths and the asterisks represent the extracted positions. It is easy to see that the estimated positions match the ground-truths well. The specific residual position errors, defined as the differences between the estimated positions and the loaded ones, are shown in Fig. 2(b). For better exhibition, the location errors along y axis are shifted by 4 pixels. It can be found that there is a pixel error at 9 positions along x axis and at 8 positions along y axis. The mean residual position errors along x axis and y axis are 0.111 and 0.099 pixels, respectively.

It is necessary to check the quality of the reconstructed image obtained by IFP with TPE position estimation. Figure 3(c) presents the TPE-IFP image recovery, while the diffraction image and IFP recovered image with the motion unknown are shown in Figs 3(a) and 3(b), respectively, as the references. We note that during IFP reconstruction procedure without the prior knowledge of motion, the translation positions used are randomly generated. Compared with the diffraction-limited image, TPE-IFP recovery presents higher resolution with more details resolvable. It also shows enhanced sharpness and contrast with respect to the one acquired by IFP with the motion unknown, which suffers from many undesirable spots. It is also noted that compared with the diffraction-limited image, IFP recovery in absence of location information still exhibits higher resolution beyond the cut-off frequency. However, the absence of the prior knowledge of motion leads to the failure of completely decoding the object and illumination pattern. As a result, the speckles left on the recovery are presented as the stains degrading the image quality. That TPE-IFP improves the resolution and eliminates the residual pattern effects on the recovery demonstrates its effectiveness in terms of pattern motion estimation.

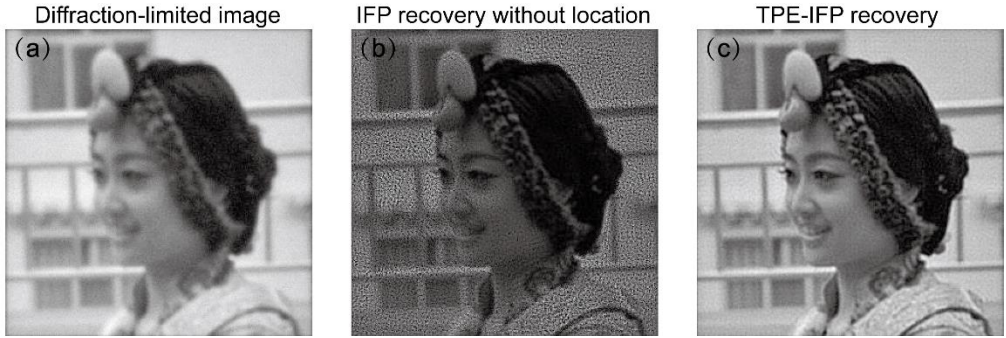

Fig. 3. Simulation results of image reconstruction. (a) The diffraction-limited image as the reference, and the images recovered by IFP (b) without and (c) with TPE.

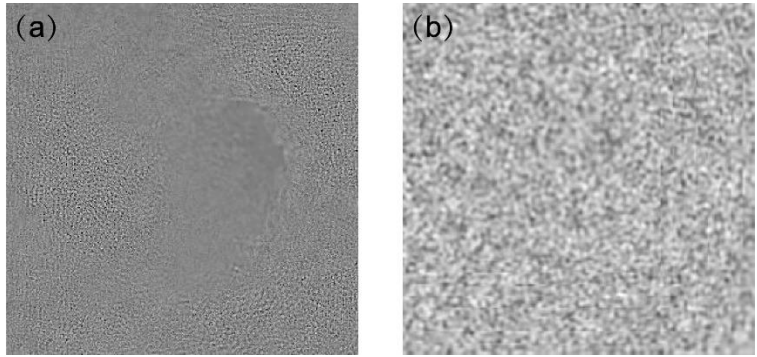

Fig. 4. Simulation results of speckle reconstruction (a) without and (b) with TPE.

It is concluded that the essential reason for the residual pattern effects on the recovery is failing to reconstruct the accurate illumination speckle during processing. Figure 4(a) presents the pattern recovered by IFP without position knowledge. As expected, it is far from the ground-truth. The pattern reconstructed by TPE-IFP is given as Fig. 4(b), which is similar to the loaded one. The results of the pattern reconstruction further indirectly provide the evidence that TPE works well.

Then we analyze the influence of additive noises on the locating accuracy using the same simulation setting. Gaussian noise is considered in this set of simulations. Different amounts of Gaussian noises are added with their variances equaling 0.5%-20% of the signal variance. The evolution of the mean position errors with noise levels is plotted as Fig. 5, from which it can be found that for both x and y position, the mean locating error initially raises and then seems converging in a fluctuating mode with the noise increasing. The largest error ranging the noise levels is less than 0.9 pixels. From this set of simulations, it can be seen that the proposed method is robust against noises.

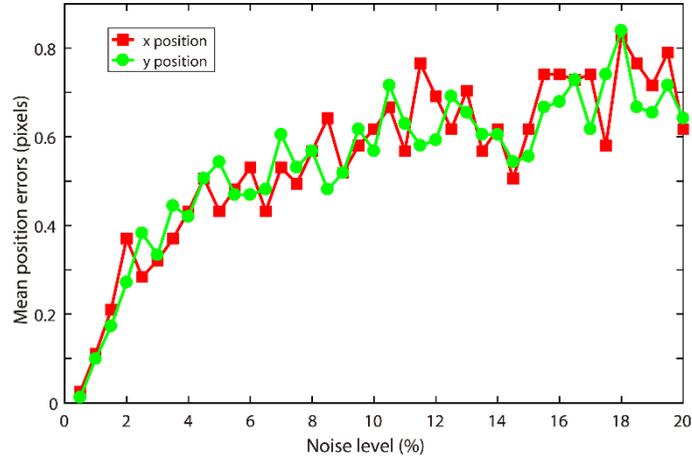

Fig. 5. Evolution of the mean position errors with respect to the noise levels.

## 4. Experimental Results

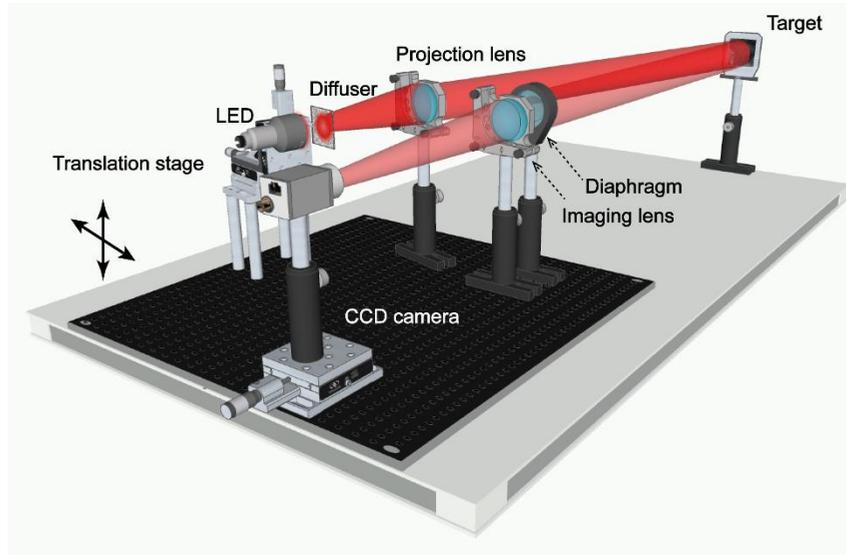

Fig. 6. The configuration of the experimental setup.

In this section, we experimentally validate the effectiveness of TPE-IFP. The experimental setup is built, as shown schematically in Fig. 6, where a 630 nm LED source and a semitransparent diffuser made by praying white paint on a slice of frosted glass are used for illumination. They are mounted on a motorized XY translation stage for spatial translation modulation. A projection lens images the diffuser illuminated by the LED onto the target, a

RMB bill, thus mixing the object and the pattern. Then an imaging system with a diaphragm against the lens collects the scattered beams, generating raw images on the CCD camera (Point Grey GS3-U3-50S5M). The diameter of the pupil is adjusted to 20 mm, and the focal length of the imaging lens is 300 mm. The pixel size of 3.45 μm is chosen to guarantee the raw images to be oversampled.

In the experiment, we scan the XY shifting stage to 64 unknown positions, and capture 64 frames of raw images corresponding to those translation locations, 4 of which are shown as samples in Figs. 7(a)-7(d). It can be seen that the pattern and the object are coupled. TPE first isolates the speckle pattern from the raw images by averaging and division, and then computes cross-correlations to extract the translation positions. Acquiring absolutely accurate positions as the references in the experiment is not as easy as doing so in the simulation. However, we can test the validity of TPE-IFP by evaluating the image and pattern reconstructions, as the simulation indicates.

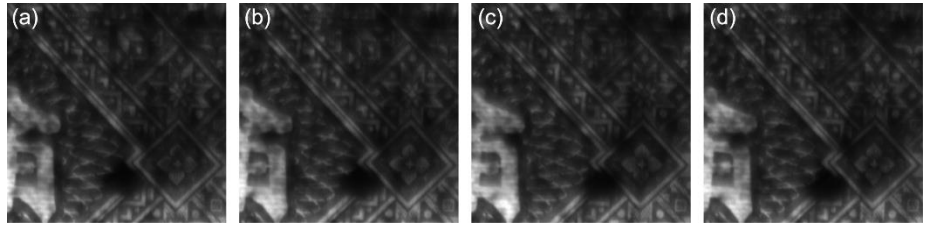

Fig. 7. Four frames of 64 raw images corresponding to unknown speckle translation locations.

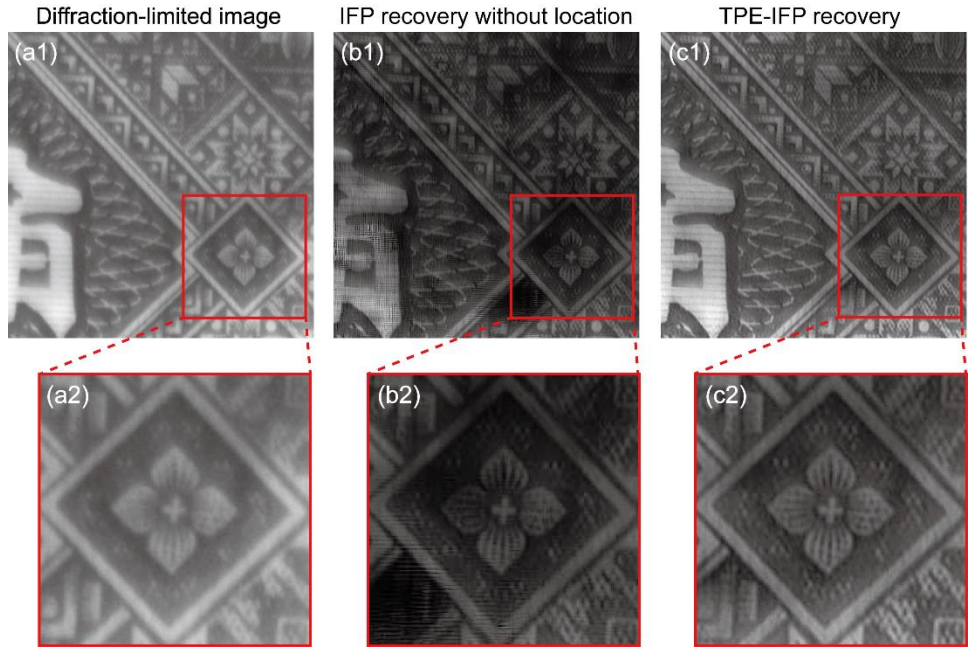

Fig. 8. Experimental results of image reconstruction. (a1) The diffraction-limited image as the reference, and the images recovered by IFP (b1) without and (c1) with TPE. (a2)-(c2) The amplified regions of interest of (a1)-(c1).

The results of image reconstruction are given in Fig. 8. Specifically, Fig. 8(c) presents the TPE-IFP image recovery, and Figs. 8(a) and 8(b) show the diffraction-limited image and IFP recovery without the motion information, respectively, playing roles as the references. The translation positions used in IFP reconstruction with the motion unknown are randomly yielded. It is obvious to see that the image reconstructed by TPE-IFP exhibits better quality with regard to the diffraction-limited one by improving the resolution. The resolution enhancement can be easily viewed from the comparison of their amplified regions, shown as Figs. 8(a2) and 8(c2). The features of horizontal and vertical bars of the petal pattern indistinct in the diffraction-limited image become resolvable in TPE-IFP recovery. Also, compared with the image recovered by the previous IFP, which suffers from residual speckles though decoding some high frequencies as shown in Fig. 8(b), the bill pattern acquired by TPE-IFP presents clearer textures and enhanced contrast. From these comparisons above, the fact that TPE-IFP presents superior image quality in absence of the translation positions demonstrates the effectiveness of TPE in motion extraction.

Similar to the simulation, the recovery using the previous IFP without translation positions exhibits residual speckles that degrade the image quality, though also presenting higher resolution as shown especially from the comparison of Figs. 8(b2) and 8(a2). Following the simulation analysis, we conclude that this case is caused by the failure of pattern reconstruction due to the lack of translation positions. The pattern recovered by IFP without position knowledge is shown as Fig. 9(a), which looks like a white board instead of a semitransparent diffuser. Luckily, our proposed TPE can bypass this problem. The pattern reconstructed by TPE-IFP is presented in Fig. 9(b), which is similar to the distribution of the semitransparent diffuser used. The experimental results of the pattern reconstruction indirectly prove that TPE is valid in translation position estimation.

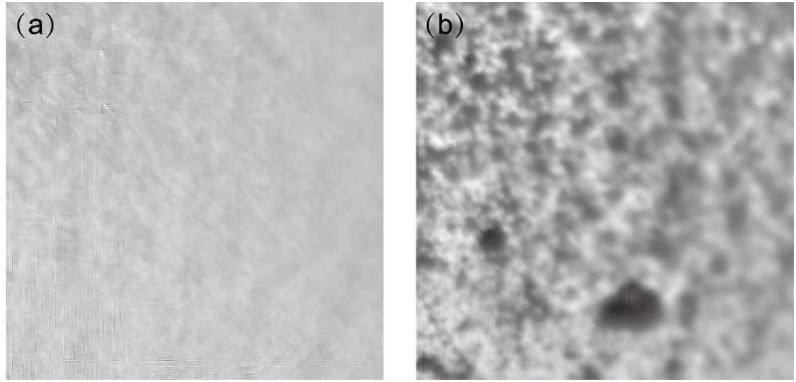

Fig. 9. Experimental results of speckle reconstruction (a) without and (b) with TPE.

## 5. Conclusion

In summary, we propose a preprocessing algorithm termed TPE for IFP supper-resolution imaging, which is capable of extracting the translation positions directly from IFP raw images. The implementation of TPE-IFP algorithm can provide us with high-quality images in the absence of the motion knowledge. In the simulations, the locating accuracy is evaluated. The mean residual position errors of 0.111 and 0.099 pixels are obtained, along x axis and y axis respectively. We also study the influence of the noise on the locating accuracy, and find that as the noise increases, the mean error initially raises and then seems converging in a fluctuating mode with the largest error less than 0.9 pixels. Then we illustrate the ability of TPE-IFP to reconstruct the high-quality image and the right illumination pattern in absence of the prior knowledge of the translation positions. In the experiment, TPE-IFP succeeds in recovering sharp super-resolution image and the right pattern speckle in the case where the illumination motion is unknown. The agreement between the simulation and experiment proves the validity of TPE in motion extracting.

With the help of TPE algorithm, TPE-IFP no longer requires the prior knowledge of the translation positions. Thus it is able to handle some complex cases where it is challenging to acquire the prior location information. By further relaxing the practical conditions, the proposed TPE may accelerate the applications of IFP in remote sensing, synthetic aperture imaging, and imaging radar. Also, as a preprocessing method irrelevant with the image reconstruction algorithm, TPE obtains the speckle movement from the raw dataset, so it might be used to determine the pattern translation positions for the similar speckle-based supper-resolution imaging techniques [8,9].

## Funding

## Disclosures

The authors declare no conflicts of interest.